\documentclass[twocolumn,showpacs,preprintnumbers,amsmath,amssymb,floatfix]{revtex4-1}

\usepackage{graphicx}
\usepackage{dcolumn}
\usepackage{bm}
\usepackage{float}
 
\begin{document}

\title{Electronic structure and excitations in oxygen 
deficient CeO$_{2-\delta}$
from DFT calculations}

 \author{T. Jarlborg$^1$, B. Barbiellini$^2$, C. Lane$^2$, 
 Yung Jui Wang$^{2,3}$, R.S. Markiewicz$^2$, 
 Zhi Liu$^3$, Zahid Hussain$^3$ and A. Bansil$^2$}

  \affiliation{
  $^1$DPMC, University of Geneva, 24 Quai Ernest-Ansermet, CH-1211 Geneva 4,
  Switzerland
  \\
  $^2$Department of Physics, Northeastern University, 
  Boston, Massachusetts 02115, USA\\
  $^3$ Advanced light Source, Lawrence Berkeley National Laboratory, 
  Berkeley, California 94720, USA
  }
 
 
\begin{abstract}
The electronic structures of supercells of CeO$_{2-\delta}$ 
have been calculated within
the Density Functional Theory (DFT).
The equilibrium properties such as lattice constants, bulk moduli
and magnetic moments are well reproduced by the 
generalized gradient approximation (GGA). 
Electronic excitations are simulated by
robust total energy calculations for constrained states
with atomic core holes or valence holes. Pristine ceria 
CeO$_2$ is found to be a non-magnetic insulator with
magnetism setting in as soon as oxygens are removed from the structure. 
In the ground state of defective ceria, the Ce-$f$
majority band resides near the Fermi level, 
but appears at about 2 eV below 
the Fermi level in  photoemission spectroscopy experiments
due to final state effects.
We also tested our computational method by calculating   
threshold energies in Ce-M$_5$ and O-K x-ray absorption 
spectroscopy and comparing theoretical 
predictions with the corresponding measurements.
Our result that $f$ electrons
reside near the Fermi level in the 
ground state of oxygen 
deficient ceria
is crucial for understanding the catalytic 
properties of 
CeO$_2$ and related materials.
\end{abstract}
 
  \pacs{71.28.+d,
        71.15.Mb
        71.15.Qe
        79.60.-i}
 
  \maketitle

\section{Introduction}
Mixed-valency cerium oxides (ceria) are 
technologically important materials 
\cite{trovarelli,esch,pirovano2007}
with remarkable properties that are useful 
for applications
in heterogeneous chemical \cite{r2} 
and electrochemical catalysis 
\cite{r3,mogensen,r5,r6}.
In chemical catalysis, ceria is used as an active support. 
Ceria at an interface catalyzes surface reactions \cite{r7,r8}, 
while the bulk material is used as an oxygen reservoir. 
In electro-catalysis, on the other hand, 
mixed ionic and electronic conductivity 
(with electrons localized around Ce) \cite{r9,r10} 
is essential for making ceria a potentially good
electrode in solid oxide fuel cells \cite{r11,r12,r13,r14}
with an outstanding electrocatalytic 
activity even without any metal
co-catalyst \cite{r15}.

Ce$^{3+}$ and oxygen vacancies are thought to be the active sites 
on ceria surfaces \cite{r16,r17,r18}
in reactions such as hydrolysis, with the
surface undergoing a Ce$^{3+}$/ Ce$^{4+}$ redox cycle during 
the complete catalytic reaction.
In the pristine CeO$_2$ compound, Ce atoms assume a +4 oxidation state, but 
the phase diagram of ceria contains a continuous range of partially reduced 
phases CeO$_{2-\delta}$ in which oxygen vacancies can be easily formed or eliminated. 
The formation of oxygen vacancies in CeO$_{2-\delta}$ 
results in changes in cerium oxidation state
similar to those implicated in the cuprates \cite{jbmb}. 
Notably, the Ce valence and defect structure in ceria
can change in response to physical parameters such as temperature, 
voltage, and oxygen partial pressure \cite{mogensen}.

The nature of the Ce active site remains not well-understood 
because studies with ceria are complicated by the fact 
that the $f$ electrons appear to be far from the Fermi 
level. The degree of participation of $f$-electrons in catalytic reactions 
\cite{watkins,galea,hutter,hellman} is therefore not clear, since
the standard theory of catalysis heavily relies on localized $d$ orbitals 
at the Fermi energy, $E_F$ \cite{norskov}. Fortunately,
the computational description of perfect CeO$_2$ 
structure is not so complicated due to the 
absence of Ce 4$f$ electrons in the insulating material 
\cite{prb_ceo2,abri}. However, in CeO$_{2-\delta}$, 
when partially filled $f$ orbitals are involved, 
the ground state predicted by the density-functional theory 
(DFT) clearly places the $f$ electrons in narrow bands 
piled at $E_F$, interacting only weakly with other electrons. 
Since the elusive $\alpha-\gamma$-transition 
in pure Ce can be described quite accurately
by temperature dependent DFT calculations 
in which vibrational, electronic and magnetic free energies are
taken into account, \cite{ce},  DFT should be expected to provide 
a reasonable description of ceria. In fact, corrections to the value of
the chemical expansion coefficient $\alpha_c$ \cite{marrocchelli}
have been explained via disorder and entropy effects
enhancing lattice expansion \cite{ce} rather than via correlations beyond the 
Generalized Gradient Approximation (GGA) \cite{gga}. 
In sharp contrast, however, signatures of $f$ bands are often found in 
spectroscopic measurements not at the Fermi energy $E_F$, 
as GGA predicts, but several eV's above 
or below  $E_F$ depending on the nature of the spectroscopy.
These shortcomings may be cured in an advanced DFT approach 
\cite{fexit}, which includes relaxation energies relevant for excitation 
of occupied and empty states in various spectroscopic probes.
These relaxation energies are of the order 
of the corrections obtained within quasi-particle 
schemes \cite{chantis,pcs66} or modeled by adding a Hubbard $U$ 
term \cite{cococcioni}. Overall, the published studies that consider  
CeO$_2$ cover the range of $U$ parameter between 2 and 8 eV,
depending on the property of interest 
\cite{andersson,ldau0,ldau1,ldau2,preda2011,spiel2011,peles2012}.
Many-body perturbation theory with $U$ \cite{jiang2009},
self-interaction corrections \cite{gerward2005},
and hybrid DFT functionals have also been considered 
\cite{graciani,gillen}. In the present study, we explore the relaxation energy 
approach \cite{fexit} to find a reasonable description 
of x-ray photoemission (XPS) as well as x-ray absorption spectroscopy (XAS) results in CeO$_{2-\delta}$.

An outline of this paper is as follows. In Sec. II, we present 
details of our electronic structure and total energy computations
for various CeO$_{x}$ supercells where $x=2-\delta$. 
Results of the calculations are presented and compared 
with relevant experimental data in Sec. III, and the conclusions 
are given in Sec. IV.

\section{Experimental setup and Method of calculation}

The photoelectron spectroscopy experiments were performed at beamline 9.3.2 at 
the Advanced Light Source (Berkeley). Detailed description of the AP-XPS 
endstation used in this study and the ceria thin films sample preparation can be 
found in Refs. \onlinecite{Ref1,Ref2}. 
The Fermi level was determined by using a gold foil. 
The binding energy was also calibrated by using the Pt 4$f$ core level. 
The Ce 4$f$ spectrum was collected at a photon energy of 270 eV 
\cite{Ref3}. In the interest of brevity, we refer to previous publications for
further details of measurements. \cite{Ref3,acscatal}

In order to determine various equilibrium properties, we used 
the pseudo-potential projected augmented wave method \cite{vasp1} 
implemented in the VASP package \cite{vasp2}
with an energy cutoff of 520 eV for the plane-wave basis set.
CeO$_2$ has a cubic fluorite lattice (Fm$\bar{3}$m) 
with four cerium and eight oxygen atoms per unit cell.
The exchange-correlation energy was computed 
using the GGA functional \cite{gga_pbe}, 
which gives a reasonable agreement
with experimental low temperature 
equilibrium volumes for CeO$_2$ and Ce$_2$O$_3$.
Andersson {\em et al.} \cite{andersson}
have pointed out that this agreement is not maintained if 
a non-zero Coulomb parameter $U$ is deployed in the GGA scheme.

To estimate the XPS and XAS relaxation effects, we have 
performed self-consistent first-principles calculations 
using the the Linear Muffin-Tin Orbital (LMTO) method \cite{lmto}
within the Local Spin Density Approximation 
(LSDA) \cite{lda} as in Ref.~\cite{fexit}
for supercells containing 4 or 8 formula units of CeO$_2$. 
The same LMTO method has been successfully applied previously
to study the effect of doping copper oxide high temperature
superconductors \cite{jbmb,bbtj}.
Here, empty spheres were inserted in the interstitial 
region opposite to the oxygen atoms, a total of 16 or 32 spheres per supercell.
Defective ceria CeO$_{x}$ configurations 
were modeled with the supercell method by considering 
two concentrations:  $x=1.75$  and $x=1.875$, in the 16 
and 32 atom supercells, respectively.\cite{AB1}
The converged self-consistent results were obtained
using a mesh of 286 or 89 $k$-points
within the irreducible Brillouin zone for the small and large 
supercells, respectively.
These calculations were made for a lattice constant $a_0$ of
5.45 \AA~ for stoichiometric CeO$_2$ 
and 5.54 \AA~ when vacancies are present.
The atomic sphere radii in the LMTO calculations are 
0.303$a_0$ for Ce, 0.230$a_0$ for O and 0.196$a_0$ 
for the empty spheres.
A precise tetrahedron method was used to determine the 
density-of-states (DOS) \cite{rath}.  

In order to calculate the XAS threshold energy, 
we start with the electronic structure 
obtained within the LMTO method.    
Our approach for modeling XAS \cite{lerch} assumes 
that the absorption is essentially a screened single-particle process. 
A step to account for many-body relaxation effects is to extract an electron 
from the core shell and add it into the valence states. 
The electronic structure computations were carried out self-consistently 
under these constrained conditions. 
After the system has relaxed, we consider the 
total energy difference between the unperturbed 
state and the relaxed state to determine 
the XAS threshold energy.

The calculation of the excitation energy 
in x-ray photoemission spectroscopy (XPS)
from the occupied Ce-$f$ state is made in the same way as 
in our earlier study of Nd$_2$CuO$_4$ \cite{fexit}.
Excitation energies for localized $f$-electrons are different from 
those for itinerant electrons, since relaxation
effects are smaller for itinerant bands.
An electron is removed at an energy lying just 
below the Fermi level from the occupied
majority state on one of the Ce atoms, and it is then spread out uniformly 
over the cell to account for a final
state at high energy. The difference in total 
energy per electron between this state and the
ground state gives the relaxation energy, 
$\Delta \epsilon$ defined in Appendix A. In particular, the final state will 
appear shifted by an amount $\Delta \epsilon$ with 
respect the Fermi level. The procedure for simulating inverse photoemission 
(or bremsstrahlung isocromat spectroscopy, BIS) is
reversed. The final state then has one $f$-electron in 
an empty Ce-$f$ state and the same
amount of opposite neutralizing charge is spread 
uniformly over the cell. These procedures assume 
large excitation energies because the compensating 
uniform charges are valid approximations 
for free electron states ignoring the lattice potential 
\cite{jn}.

\section{Results}
\subsection{Ground-state properties}
\begin{figure}[h]
  \begin{center}
  \includegraphics[width=8cm,height=6cm]{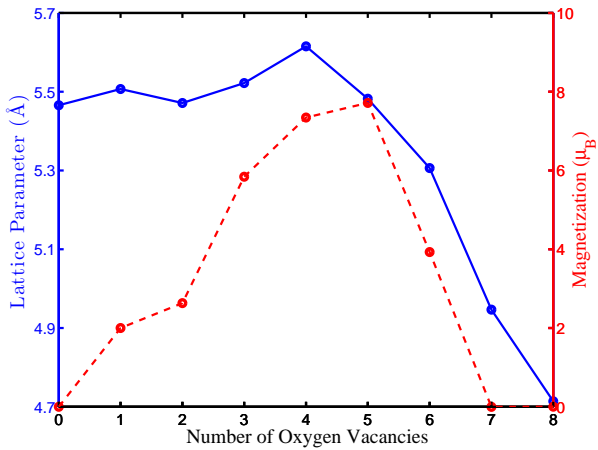}
  \end{center}
  \caption{(Color online) Lattice constant (points connected by solid line) 
and magnetic moment (points connected by dashed line) 
results of the VASP based calculations on a Ce$_4$O$_{8-N}$ cell
as function of the number $N$ of oxygen vacancies. 
Both volume and atomic positions were relaxed while
keeping the cubic symmetry. For each $N$ the configuration with the lowest energy was chosen.} 
  \label{figvasp}
  \end{figure}

The VASP calculation on a Ce$_4$O$_8$ cell gives an equilibrium lattice constant  
$a_0=5.466$ \AA~ and a bulk modulus $B_0=198.9$ GPa, which compare well with the corresponding 
low-temperature experimental values $a_0=5.41$ \AA~ \cite{eyring}
$B_0=204$ GPa \cite{nakajima}. By removing an oxygen atom and by letting 
the volume and the atomic positions relax,
the lattice constant was found to expand to $a=5.507$ \AA~, 
the bulk modulus reduced to $B=155.9$ GPa, 
and the total spin magnetic moment was 2 $\mu_B$.
This predicted ferromagnetic structure is consistent 
with experiments \cite{fernandes1,fernandes2,chen}
and other first principles studies \cite{han}. 
We note that by removing all the oxygen atoms, one recovers 
the fcc-phase of Ce, and the 
calculated bulk modulus is only 51.4 GPa for 
the non-magnetic $\alpha$ phase 
at an equilibrium lattice constant of 4.714\AA.
Our results for Ce$_4$O$_{8-N}$ 
as a function of the number $N$ of oxygen vacancies
are summarized in Fig.~\ref{figvasp}.
The calculated chemical expansion coefficient 
of CeO$_{2-\delta}$ is given by
$\alpha_c=(a-a_0)/(a_0 \delta)$.
However, our results for O-vacancies in the
small unit cell Ce$_4$O$_{8-N}$ 
show that $\alpha_c$ is not at all linear. In fact,
between $N=0$ and $N=1$, $\alpha_c$ is $0.03$, while 
between $N=3$ and $N=4$, 
$\alpha_c=0.07$, which is slightly below the experimental value  
$\alpha_c=0.08-0.1$ \cite{marrocchelli}.
Part of the anomalous behavior of $\alpha_c=0.07$
can be explained in term of oxygen vacancy ordering leading to  
lattice contractions opposite to the chemical expansion\cite{kuru}.
The GGA calculation for one vacancy in a large $2\times 2 \times 2$
supercell with $95$ atoms gives a larger dilute vacancy limit ($\delta \rightarrow 0$) $\alpha_c=0.05$ as shown by Marrocchelli {\em et al.}.  
Therefore, calculated values of $\alpha_c$ are not easy to compare with experiment within large ranges of $T$ and $\delta$.
The discrepancy between DFT and the experiment 
could be explained by disorder and entropy effects leading to larger 
high-temperature equilibrium volumes \cite{ce}. 
The electronic entropy will increase the 
lattice constant of oxygen-deficient ceria, since the 
DOS near the Fermi level is higher at large volume and large $\delta$, 
although this effect is rather weak.  
Lattice disorder is caused both by zero-point
motion \cite{gga} and thermal vibrations \cite{fesi},
which produce a pressure given by $\-d\omega/dV$, 
where the phonon frequency $\omega$ is proportional to
$\sqrt{a B}$. Also the effect of spin and orbital magnetic fluctuations 
usually produce lattice expansions as shown in Ref.~\cite{ce}. 
The relevant temperature range for catalytic applications is rather high 
(i.e. $500-700$ $^{\circ}$C) \cite{Ref3} and the effects 
from lattice vibrations on magnetic fluctuations 
in this range make the total entropy balance complicated.
 \begin{figure}[h]
  \begin{center}
  \includegraphics[width=8cm,height=6cm]{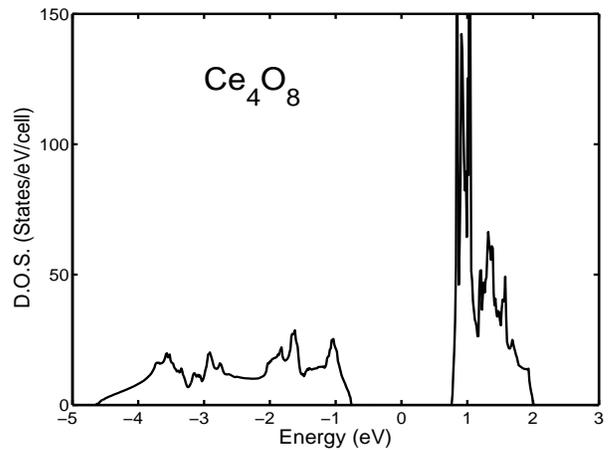}
  \end{center}
  \caption{Total DOS for Ce$_4$O$_8$ calculated from 286 $k$-points.
   The Fermi level is at zero.} 
  \label{fig16nm}
  \end{figure}

\begin{figure}[h]
  \begin{center}
  \includegraphics[width=8cm,height=6cm]{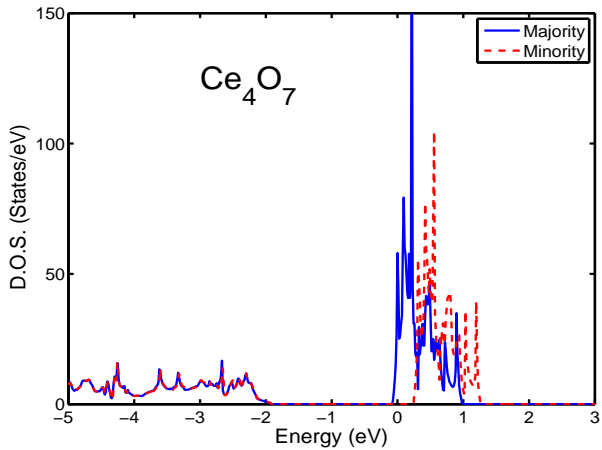}
  \end{center}
  \caption{(Color online) Total spin-polarized DOS for 
  Ce$_4$O$_7$ calculated from 286 $k$-points using the LSDA.
   The Fermi level is at zero. } 
  \label{fig16fm}
  \end{figure}

\begin{figure}[h]
  \begin{center}
  \includegraphics[width=8cm,height=7cm]{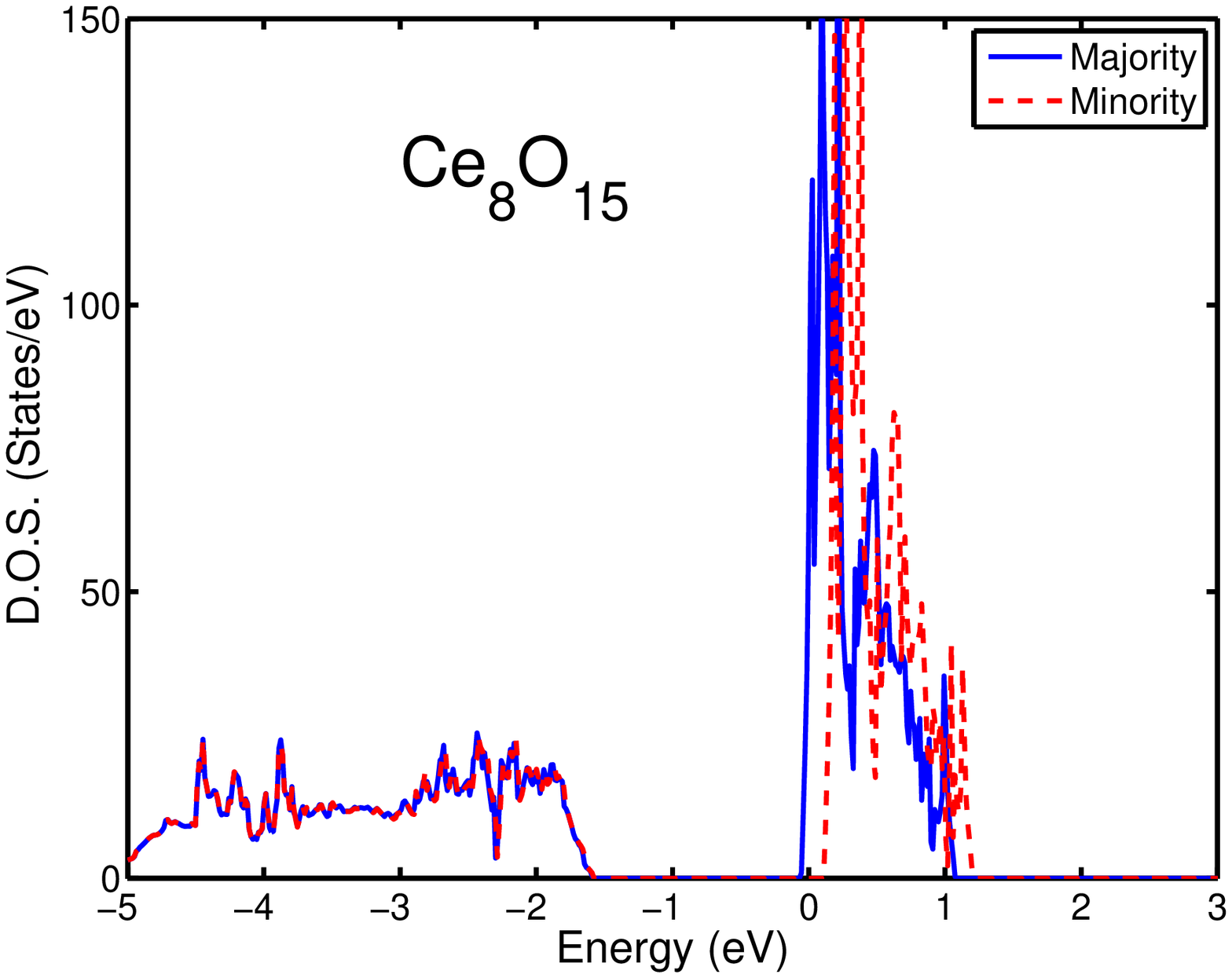}
  \end{center}
  \caption{(Color online) Total spin-polarized DOS for 
   Ce$_8$O$_{15}$ calculated from 89 $k$-points using the LSDA.
   The Fermi level is at zero. } 
  \label{fig32fm}
  \end{figure}

The LMTO electronic structure of CeO$_2$ 
is found to be non-magnetic and
insulating as shown in Fig. \ref{fig16nm}.
The distance between the valence band 
and the Ce 4$f$ edge is about 1.6 eV,
to be compared with 3 eV 
in experiments \cite{prb_ceo2,delley}. 
Ferromagnetism (FM) is not expected because 
of the absence of occupied Ce 4$f$ electrons. 
An oxygen atom has 4 valence $p$ electrons. 
But all 3 O-2$p$ bands are below $E_F$ and 
can harbor 6 electrons (2 spins in each band). 
Therefore, each removal of an O-atom
removes 3 occupied $p$-bands, but since the system has 
only 4 fewer electrons, this
means that $E_F$ will rise, and one more band will 
be occupied to account for the two additional electrons. Thus,
an oxygen vacancy introduces partially filled Ce $f$ states and the FM ordering sets in because the high DOS of Ce-$f$ states produces a Stoner exchange splitting. The calculated moment is 0.52 $\mu_B$ per Ce atom in CeO$_{1.75}$ and the FM state has a lower total energy ($E_{tot}$) than the non-magnetic state by 0.14 eV per formula unit. The induced moment per oxygen is negative, about 0.01 $\mu_B$, and the total moment per Ce$_4$O$_7$ cell is exactly 2.0 $\mu_B$, or 0.50 $\mu_B$ per CeO$_{1.75}$ unit. The FM state is half-metallic with no minority bands at $E_F$, and as expected from the qualitative discussion above,
the charge transfer to the majority states is exactly 2 spin states per cell,
see Figs. \ref{fig16fm} and \ref{fig32fm}.
Consequently, there also are exactly 2 more majority states than minority states
in the calculation for the 32-site cell Ce$_8$O$_{15}$, 
which corresponds to CeO$_{1.875}$. Here there are two types of Ce sites, the 4 closest to the O-vacancy, Ce$_v$, has each a moment of 0.33 $\mu_B$ and the 4 towards the interior, Ce$_i$, have 0.19 $\mu_B$ each \cite{charge}. Together with the small negative moments on the oxygen this gives 
exactly 2.0 $\mu_B$ per Ce$_8$O$_{15}$ cell, or 
0.25 $\mu_B$ per CeO$_{1.875}$ unit. In other words, the removal of one
oxygen atom gives rise to a spin magnetic moment of $2 \mu_B$
in the dilute vacancy limit ($\delta \rightarrow 0$). 

\subsection{Excited-state properties}
Table~\ref{table1} compares measured positions of 
several Kohn-Sham core energy levels with the corresponding 
calculated XAS threshold energies.
Clearly, the total energy calculations give XAS threshold 
energies in much better agreement 
with experiments \cite{yagci,melcher,chen}
compared to just taking the Kohn-Sham energy of the core level relative 
to $E_F$. Note that we are not looking for absolute values, 
but rather for relative differences between
FM and non-magnetic configurations with or without vacancies.
In particular, the value of the Ce M-edge threshold is reduced by about 
1 eV for an oxygen vacancy in the small supercell Ce$_4$O$_7$
and this value is consistent with the displacements
of the Ce$_2$O$_3$ \cite{melcher} and  
of the metallic Ce \cite{yagci} threshold energies toward lower values.
In the large super-cell, Ce$_8$O$_{15}$, the Ce-3$d$ Kohn-Sham core levels 
differ by 0.44 eV between the two types of Ce sites.
The threshold of oxygen K-edge position seems less affected 
by the formation of O vacancies in ceria in agreement with 
measurements performed in oxygen-deficient CeO$_2$ nanoparticles \cite{chen}.

\begin{table}[ht]
\caption{\label{table1}
Core level Kohn-Sham energies $(\epsilon)$ 
and the calculated threshold energies $(E_{XAS})$ 
for x-ray absorption for the upper spin-orbit 
Ce-d$_{5/2}$ state (M$_5$-level) 
and O-1$s$ state (K-level), all in units of eV.
The spin polarized potentials $V^\sigma$ produced 
by magnetic valence (non-relativistic) electrons
affect the position of the M$_5$-level.
($E_F$ for the insulating Ce$_4$O$_8$ is assumed to be in the middle of 
the gap of 1.54 eV.)
The calculated Ce-3$d$ spin-orbit splitting is 18.82 eV.}
  \vskip 2mm
  \begin{center}
  \begin{tabular}{l c c c}
  \hline
      core hole  & $(E_F-\epsilon)$  & $E_{XAS}$  & experiment\\

  \hline \hline

Ce-3$d_{5/2}$ in Ce$_4$O$_8$    & 856.2 &  871.6 & 877 Ref.~\cite{yagci} \\
O-1$s_{1/2}$ in Ce$_4$O$_8$    & 501.0 &  523.7 & 528 Ref.~\cite{chen}\\
Ce-3$d_{5/2}$-maj. in Ce$_4$O$_7$    & 856.0 &  870.8 & \\
Ce-3$d_{5/2}$-min. in Ce$_4$O$_7$    & 856.0 &  870.3 & \\
O-1$s_{1/2}$ in Ce$_4$O$_7$    & 502.0 &  523.5 &\\
  \hline
  \end{tabular}
  \end{center}
  \end{table}

The results for the XPS and BIS excitation energy per electron 
given in Table \ref{table2} show trends similar to those given 
in Ref.~\cite{fexit} for Nd$_{2-x}$Ce$_x$CuO$_4$.
In particular, the relaxation corrections for defective ceria 
split the single $f$ peak in LSDA to an occupied 
and an unoccupied $f$ band. 
The former falls below and the latter lies above $E_F$. 
Our XPS calculations predict that the $f$ occupied peak appears at
about 1 eV below the Fermi level while
experimentally the position of this peak seems to be even lower.
Typical {\em in situ or in operando} Ce 4$f$ and Ce 3$d$ x-ray photoelectron 
spectra of a ceria electrode are shown in Fig. 3a of Ref.~\cite{Ref3}.
Figure 7 of Ref.~\cite{acscatal} provides the 
spatially resolved spectral image of the Ce 4$f$ valence band, allowing visualization of 
regions of electrochemical activity in a 
ceria electrode. In fact, the presence of Ce$^{3+}$ species is revealed 
by the intensity of the Ce 4$f$ occupied peak at about 2 eV binding energy
as already demonstrated by {\em ex situ} results~\cite{exp_xps}.
The Ce 3$d$ XPS core-level spectra display 
different final-state populations of 4$f$, 
which lead to the peak splitting shown 
in Fig. 3b of Ref.~\cite{Ref3}.
The final state effects lead to 
an upward shift of the lowest binding Ce-3$d_{5/2}$ 
peak attributed to the Ce$^{3+}$ state 
[see also Ce 3$d$ XPS spectra in Fig.1 of Ref.~\cite{exp_xps}],
which is consistent with the shift of the calculated 
Ce 3$d_{5/2}$ XAS threshold \cite{AB2} given in Table \ref{table1}.

For the smaller unit cell, Ce$_4$O$_7$, where all the Ce atoms are equivalent, 
we find that the occupied $f$ band shifts only by $-0.7$ eV,
indicating that the degree of localization of the $f$ orbital
plays an important role in the value of the energy shift.
If we take the atomic positions relaxed by VASP,
the energy shift becomes $-0.8$ eV.
Therefore, the correction due to atomic relaxation is of the order of $0.1$ eV.
Calculations were also made for the large cell, Ce$_8$O$_{15}$, where 
both Ce$_v$ and Ce$_i$ give the same position for the $f$ occupied peak 
at about 1 eV below $E_F$ \cite{footnote1}.
This is consistent with the observation that 
the Ce 4$f$ binding energy does not change much with 
the vacancy concentration, but that different contributions from different sites
lead to some broadening. Therefore, the excess electrons left behind by the removal of neutral oxygen atoms produce occupied $f$ states with practically the same 2 eV binding energy. The reason for the underestimation of the theoretical XPS relaxation is not known, but several mechanisms may be involved.
As shown above, part of the correction is due to lattice relaxation near the vacancies. Other possible modifications of the Ce potential would result at surface sites or due to atomic vibrations. Interestingly, the calculated shift for the $f$ states in Nd$_{2-x}$Ce$_x$CuO$_4$ \cite{fexit} has also been found to be about 2 eV. It is also possible that the approximation of a completely 
delocalized excited state is less appropriate for Ce at these energies.

\begin{table}[ht]
\caption{\label{table2}
 Calculated relaxation, $\Delta \epsilon$, of Ce-$f$ states in XPS and BIS in eV.}
  \vskip 2mm
  \begin{center}
  \begin{tabular}{l c c c}
  \hline
     excitation  & Ce in Ce$_4$O$_7$  & Ce$_v$ in Ce$_8$O$_{15}$ & Ce$_i$ in Ce$_8$O$_{15}$  \\

  \hline \hline

XPS    & -0.7 &  -1.0 & -1.1 \\
BIS    & 1.0 &  1.1 & 1.0 \\

  \hline
  \end{tabular}
  \end{center}
  \end{table}
  
\section{Conclusions}

We have obtained electronic structures of supercells of CeO$_{2-\delta}$ within 
the framework of the DFT. The experimental equilibrium lattice constants, 
bulk moduli and magnetic moments are well reproduced by the 
generalized gradient approximation (GGA) without 
the need to introduce a large Coulomb parameter $U$. 
The computed value of lattice chemical expansion $\alpha_c$ as a function of O-vacancy concentration
is not linear for $\delta$ ranging from 0 to 1. 
Pristine CeO$_2$ is found to be a non-magnetic insulator with
magnetism setting in as soon as oxygens are removed from the structure. 
Excitation properties are simulated via constrained total energy calculations, 
and Ce-M and O-K edge x-ray absorption threshold energies are discussed. 
Our study shows that the way the ground state is probed by different 
spectroscopies can be modified significantly through final state effects. \cite{fexit} 
In particular, these relaxation effects yield
a renormalization of $f$-levels away from the
Fermi Level for electron excitation spectroscopies. 
Our result that $f$ electrons reside near the Fermi level in the ground state of oxygen 
deficient ceria is crucial for understanding catalytic 
properties of CeO$_2$ and related materials.\cite{norskov}.

\acknowledgments  
We acknowledge useful discussions with Dario Marrocchelli.
The work at Northeastern University is supported by the 
US Department of Energy (USDOE) Contract No. DE-FG02-07ER46352.
The Advanced Light Source is supported by the Director, 
Office of Science, Office of Basic Energy Sciences, 
of the USDOE under Contract No. DE-AC02-05CH11231.
We benefited from computer time from Northeastern 
University's Advanced Scientific Computation Center (ASCC) 
and USDOEÕs NERSC supercomputing center.
 
\appendix
\section{Details of Constrained DFT Computations}

In the XPS final state, one whole electron is 
transferred from the ground-state to a homogeneous 
plane-wave single-particle state. We can simulate this process by creating 
a hole obtained by removing states from the local DOS (LDOS)
over a narrow energy window $[E_1,E_F^*]$, where $E_F^*$ is the Fermi energy 
in the excited state and $E_1 \le E_F^* $ 
is a cut-off energy. 
The electronic density $\rho_h(r )$ associated with the hole at the site $t$ is
 \begin{eqnarray}
 \rho_h(r )= \sum_{\ell}\int_{E_1}^{E_F^*}N_{t,\ell}^*(E)R^2_{t,\ell}(E,r)dE 
 \label{rhoh}
 \end{eqnarray}
Here, $N_{t,\ell}^*(E)$ is the self-consistent LDOS and $R_{t,\ell}(E,r)$ 
is the radial wave function component at the site $t$ with angular momentum $\ell$. 
The total electron density $\rho^*(r)$ for the excited state thus is
\begin{eqnarray}
 \rho^*(r )=\sum_{t',\ell}\int_{-\infty}^{E_F^*} N_{t',\ell}^*(E)R^2_{t',\ell}(E,r)dE
 \nonumber \\
 - \rho_h(r ) +
 \frac{1}{\Omega},
 \label{rho}
 \end{eqnarray}
where the last term in Eq.~\ref{rho} imposes charge neutrality within 
the simulation cell of volume $\Omega$.

The charge density in the final state $\rho^*(r )$ allows us to
determine the total energy $E^*$ of the excited state. 
The electrostatic and exchange-correlation contributions 
are evaluated straightforwardly from $\rho^*(r )$. 
The kinetic energy corresponding to the
excited charge density does not involve a single energy level of the solid 
but rather a group of states and it can be calculated
using the standard expression given by Janak \cite{janak}, which involves 
the Kohn-Sham energy average
\begin{eqnarray}
 \epsilon^*= \sum_{\ell}\int_{E_1}^{E_F^*}N_{t,\ell}^*(E)EdE. 
\label{epsprime}
\end{eqnarray}
In this way, the total energy $E^*$ can be obtained in terms of the Kohn-Sham 
energy average $\epsilon^*$ and the hole density $\rho_h$
exactly as in the XAS threshold energy calculations within the $\Delta$ 
self-consistent-field ($\Delta$SCF) method \cite{lerch,hedin}. 
Finally, the energy ${\cal E}$ of the XPS photoelectron 
is given by \cite{fujikawa}
\begin{equation} 
{\cal E}= \hbar\omega + (E_0-E^*)-E^*_F,
\label{Eq_scf}
\end{equation}
where $\hbar \omega$ is the photon energy, $E_0$ is the ground state energy, 
and $E^*_F$ is the Fermi level for the excited state. If KoopmansÕ theorem applies 
\begin{equation}
{\cal E}= \hbar\omega + \epsilon - E_F,
\label{Eq_koop}
\end{equation}
where the Kohn-Sham energy average
\begin{eqnarray}
\epsilon= \sum_{\ell}\int_{E_1}^{E_F}N_{t,\ell}(E)EdE 
\label{eps}
\end{eqnarray}
is calculated in the ground state. The difference between
Eq.~\ref{Eq_scf} and Eq.~\ref{Eq_koop}
defines the relaxation energy $\Delta \epsilon$.
DFT is expected to give a reasonable description
of the energy difference $\Delta \epsilon$ 
involved in the photoemission process\cite{dabo}.

In order to focus on the excitation corresponding
to a given DOS peak, in actual computations, we considered a narrow 
energy interval $[E_{\beta},E_F^*]$ containing 
a fraction $\beta$ of an electron per site $t$, 
and renormalized the result to obtain $\Delta \epsilon(\beta)/\beta$
to account for a whole electron 
involved in the XPS process \cite{footnote1}. 
We have performed test computations using a range of $\beta$ values 
and found that the relaxation energy is very insensitive to the value of $\beta$ used, 
which is also anticipated from the analysis of Ref. \cite{ley}. 
Note that placing the hole on only one of the sites in the unit 
cell is an approximation for removing a band electron in that it 
neglects 
the overlap of the wave function with neighboring sites.
This however is expected to be a reasonable approximation for localized 
$f$-electrons of interest here. To check this point, 
we computed the excitation energy self-consistently in Ce$_4$O$_7$ 
where we removed 1/4th of an $f$-electron from each of the 4 Ce atoms 
in the unit cell, i.e. a total of one electron from the unit cell.
The value of the relaxation energy so obtained was 0.75 eV compared to 0.7 eV 
shown in Table II, which 
is within the intrinsic error of 0.05 eV in our total energy computations.


\begin{thebibliography}{99}
\bibitem{charge} The valence charge is 0.05 electrons larger 
on the Ce with small moment compared to the 
other Ce site.
\bibitem{footnote1}
This does not mean that we are exciting a fraction of an electron $\beta$ since each Bloch electron is 
spread over a macroscopic number of unit cells in the crystal, and its spectral weight on any atom is 
negligibly small. All results in Table II are based on using 
$\beta$ in between 0.125 and 0.45.
\end{thebibliography}
\end{document}